\documentclass{article}

\usepackage{amssymb,amsfonts,amsmath}
\usepackage{cite,enumerate,float}
\usepackage{color}
\usepackage{tikz,tcolorbox}

\usepackage{xcolor}
\definecolor{mintcream}{rgb}{0.96, 1.0, 0.98}
\definecolor{champagne}{rgb}{0.97, 0.91, 0.81}
\definecolor{bubblegum}{rgb}{0.99, 0.76, 0.8}
	\definecolor{airforceblue}{rgb}{0.36, 0.54, 0.66}
\usepackage[pdftex,unicode,
colorlinks=true,
linkcolor = blue,
citecolor = cyan,
urlcolor = airforceblue]{hyperref}
\usepackage{url}
\usepackage{mathtools,array}
\usepackage{comment}
\usepackage{arydshln}

\usetikzlibrary{arrows,snakes,backgrounds}

\def\be{\begin{eqnarray}}
\def\ee{\end{eqnarray}}
\def\nn{\nonumber}

\def\p{\partial}

\definecolor{red}{rgb}{1,0,0}
\definecolor{orange}{rgb}{1,0.5,0}
\definecolor{violet}{rgb}{0.7,0,1}

\hoffset= - 1.0in

\begin{document}

\hfill 

\hfill   MIPT/TH-17/23

\hfill IITP/TH-16/23

\hfill ITEP/TH-22/23

\bigskip

\centerline{\Large{
	On factorization hierarchy of equations for banana Feynman amplitudes
}}
\bigskip

\centerline{	V. Mishnyakov$^{a,b,c,}$\footnote{mishnyakovvv@gmail.com},
	A. Morozov$^{a,c,d,}$\footnote{morozov@itep.ru},	M.Reva$^{a,c,}$\footnote{reva.ma@phystech.edu}}

\bigskip

\begin{center}
	$^a$ {\small {\it MIPT, Dolgoprudny, 141701, Russia}}\\
	$^b$ {\small {\it Lebedev Physics Institute, Moscow 119991, Russia}}\\
	$^c$ {\small {\it NRC ``Kurchatov Institute", 123182, Moscow, Russia}}\\
	$^d$ {\small {\it Institute for Information Transmission Problems, Moscow 127994, Russia}}\\
\end{center}

\bigskip

\centerline{ABSTRACT}

\bigskip

{\small
We present a review of the relations between various equations for maximal cut banana Feynman diagrams, i.e. amplitudes with propagators substituted with $\delta$-functions. We consider both equal and generic masses. There are three types of equation to consider: those in coordinate space, their Fourier transform and Picard-Fuchs equations originating from the parametric representation. First we review the properties of the corresponding differential operators themselves, mainly their factorization properties at the equal mass locus and their form at special values of the dimension. Then we study the relation between the Fourier transform of the coordinate space equations and the Picard-Fuchs equations and show that they are related by factorization as well. The equations in question are the counterparts of the Virasoro constraints in the
much-better studied theory of eigenvalue matrix models and are the first step
towards building a full-fledged theory of Feynman integrals,
which will reveal their hidden integrable structure.
}

\bigskip

\bigskip

\section{Introduction}   

Lifting the theory of Feynman diagrams  \cite{Weinzierl:2022eaz,Duhr:2014woa,Vanhove:2018mto,Rella:2020ivo,Brown:2015fyf,Loebbert:2022nfu,Klausen:2023gui} to the level
achieved in the study of the eigenvalue matrix models \cite{morozov1994integrability,morozov1992string,Mironov:1993wi},
is one of the current tasks in theoretical physics.
For a bird-fly view of the subject and its connection to the
celebrated Connes-Kreimer hidden symmetry \cite{Connes:1999zw} of Feynman graphs
see \cite{Gerasimov:2000pr}.
Still, at practical level the current stage of development is far more modest:
majority of papers in the field \cite{Lairez:2022zkj,Muller-Stach:2012tgj,Kotikov:2021tai} are targeted at the study of Ward identities,
i.e. differential equations on individual Feynman amplitudes, considered as functions
of external parameters.
This is supposed to be in parallel with the Virasoro-constraint stage \cite{morozov1994integrability,Mironov:1993wi}
of the matrix model studies.

Moreover, the study of these equations is focused mostly on the
algebro-geometric interpretation, which considers them as Picard-Fuchs (PF) equations
for integral (periods) dependence of external parameters (moduli) \cite{Muller-Stach:2012tgj,Lairez:2022zkj, Mishnyakov:2022uer}.
One natural family of equations is related to the nice families of certain toric Calabi-Yau spaces \cite{Vanhove:2018mto,Bonisch:2021yfw} -- which however does not cover Feynman integrals in all space-time dimensions.

In \cite{Mishnyakov:2023wpd} it was suggested to approach the problem from a different side,
and study the families of Feynman graphs {\it per se}, starting from the
simplest class of banana diagrams. The first simplification that this case brings on is that, being two-point functions, these Feynman integrals depend on a single
parameter $x^2$ in the coordinate or $t=p^2$ in the momentum space representations. The second and important simplification from the point of view of \cite{Mishnyakov:2023wpd} is that in coordinate space the  banana amplitude with $n-1$ loops is just the product of
$n$ Green functions. It therefore satisfies a conceptually trivial equation
\begin{equation}
	\hat{\cal E}^{(n)}_{x^2} B_n(x)=0
\end{equation}
implied by the equation $(\Box+m^2)G=\delta(\vec x)$ on a single Green function. For further simplification we omit the $\delta$-function at the r.h.s. and
consider the propagators given by the solution of the "homogeneous" equation. The operator $\hat{\cal E}^{(n)}_{x^2} $ is nothing but a tensor product of the basic Klein-Gordon operator. The important property of this equation is the its solutions are \emph{only} the $n$-fold products of propagators, therefore the banana amplitudes $B_n(x)$ are uniquely defined as the space of solutions to these equations.  While banana amplitude is trivial in position space, it is given by for more complicated $n-1$ fold integrals in momentum space.  On the other hand, being a Fourier transform, they should satisfy equations given by Fourier transformed operators  $\hat{E}^{(n)}_{t} $:
\begin{equation}
	\hat{\cal E}^{(n)}_{x^2} \xrightarrow[]{\quad \text{Fourier transform} \quad } \hat{E}^{(n)}_{t}
\end{equation}
 The passing to ''homogeneous'' equations implies the substitution of
Feynman propagator  by its cut (imaginary part):
\begin{equation}
	\frac{1}{t-m^2} \  \longrightarrow  \  \delta(t-m^2)
\end{equation}
The corresponding amplitudes with all propagators replaced by $\delta$-functions are called maximal cuts. On the other hand these maximal cuts of banana integrals correspond to taking closed contour integration in the parametric integrals and hence to  {\it periods} and associated Picard-Fuchs equations
$\widehat{PF}^{(n)}_{t} B_n(t) = 0$. Here we denote by $\widehat{PF}^{(n)}_{t} $ the operator derived from the parametric integral. One should naturally wonder, whether the two differential equations given by $\hat{E}_t^{(n)}$ and $\widehat{PF}^{(n)}_t$ are related.

To set up the conjecture we notice that it is quite easy to understand that the coordinate space operator $\hat{\cal E}^{(n)}_{x^2} $
is a differential operator of exponential order -- $2^n$ in $\p_x$ when all masses are different. This happens already in the trivial case of $D=1$ and is a manifestation of the simple fact that there are $2^n$ independent $n$-fold products of the original two solutions to the Klein-Gordon equation. Counting the order of the differential operator in momentum space requires also keeping track of the dependence of $\hat{\cal E}^{(n)}_{x^2}$ on $x^2$ instead of just the derivatives. Even this appears to be a complicated problem, and we can currently only conjecture, that the order of $\hat{E}^{(n)}_{t}$  in $\p_t$ also grows exponentially with $n$. We will discuss the known cases below.
Similarly one could ask, what is the order of $\widehat{PF}^{(n)}_{t}$?  This seems to be an even more sophisticated question and currently known examples do not allow to differentiate between supposedly polynomial or exponential growth in $n$.

A general phenomenon that one encounters is that all of the discussed operators tend to factorize, often dramatically dropping the order at coinciding masses or certain critical dimension, such as $D=2$ in momentum space. For example, for equal masses the position space operator $\hat{\cal E}^{(n)}_{x^2}$ and both  $\hat{E}^{(n)}_{t}$ and $\widehat{PF}^{(n)}_{t}$ have order $n+1$ and $n-1$ respectively. This phenomenon provides a plethora of differential operators even at given $n$ and hence a certain \emph{factorization hierarchy} is present.

In view of the above discussion and based on the studied examples we conjecture that the relation between the Fourier transformed $\hat{E}_t^{(n)}$ and the $\widehat{PF}^{(n)}_{t} $ is also that of \emph{factorization}:
\be
\hat{E}_t^{(n)} = \hat {\bf B}\cdot \widehat{PF}^{(n)}
\label{factEBPF}
\ee
Where $\hat{\bf B}$ is some differential operator. Factorization of polynomials in the case of a single variable
is fully described by the fundamental theorem of algebra - any polynomial is factorizes over an algebraically closed field, while factorization, say, over $\mathbb{Q}$ is not always possible, however, various methods and criteria are known. It is far less simple for differential operators -- even in a single variable \cite{VANDERHOEVEN2007236}. Statements like Polya's factorization \cite{polya1922mean,Etingof1997FactorizationOD} are known for rather general operators:
\begin{equation}
	\sum_{i=0}^n a_i(t) \dfrac{\partial^i}{\partial t^i}  = \prod_{i=1}^{n} \left( \dfrac{\partial}{\partial t} +b_i (t)\right)
\end{equation}
However, even if $a_i(t)$ are polynomial (rational) coefficients, the functions $b_i(t)$ can be complicated multivalued functions. Here we will call an operator factorizable if it can be factorized into other operators with polynomial (rational) coefficients. However criteria for existence of factorization are rather involved and algorithms are known only in special cases .  

Picard-Fuchs operators are reasonably studied in some cases, like Calabi-Yau operators \cite{bogner2013algebraic,almkvist2010art,almkvist2013class,van2017calabi},
which has only accidental intersections with banana Feynman amplitudes. However even in this case no factorization simplifications are presented.
Additional complication is that operator $\hat {\bf B}$ in  (\ref{factEBPF})
is not necessarily with polynomial coefficients --
in general they are {\it rational} functions of $t$, and a better formulation
of factorization is
\be
\boxed{
P\cdot \hat{E}^{(n)}_t = \hat { B}\cdot \widehat{PF}^{(n)}
\label{factPEBPF}
}
\ee
with a polynomial $P(t)$ and an operator $\hat B(t,\p_t)$ with polynomial coefficients.

 Our goal in this overview paper is to illuminate the current status of the relation between the operators $\hat{E}^{(n)}_{t}$ and $\widehat{PF}^{(n)}_{t} $ and, furthermore, their factorization properties at special points. Not much is known about this factorization hierarchy even for banana diagrams, which is why we try to structure the known examples and conjecture some general properties to set up the stage for further, more technical, works. It is a technically involved problem, and an adequate language,
which would make the answers as simple as the questions, is still to be found.
For technical details behind this overview see \cite{MMRSlarge}.

\section{Equations $\hat{\cal E}^{(n)}$ in coordinate space}

\subsection{The basic case of $D=1$}
Cut propagators are solutions of the equation
$(\hat\Box + m^2)G_m = 0$, which is the simplest for $D=1$:  then $G_{\pm m} = e^{\pm imx}$ since $\Box =  \partial_x^2$.
A banana Feynman diagram in coordinate space is a product of $n$ such functions:
\begin{equation}
	B_n(x) \Big|_{D=1}= \prod_{i=1}^n e^{\pm m_i x}
\end{equation}
with no integration.
These functions can clearly be uniquely defined as the  set of solutions of the equation
\be
\prod_{\pm} \Big(\hat\Box + (m_1\pm m_2 \pm\ldots \pm m_n)^2\Big) \prod_{i=1}^n G_{\pm m_i} = 0
\label{coorddifmD1}
\ee
where the product is over all the possible combinations of $+$ and $-$ between the masses. This equation is of the order $2^{n-1}$ in $\hat \Box := \partial^\mu\ \partial_\mu \big|_{D=1}=\p_x^2$ operator.
For the equal mass case some of the alternate sums of the masses become equal. The simplest example is $m_1 - m_2 $ and $m_2-m_1$  for $n=2$ which both vanish at equal masses.
Hence the number of independent banana functions is now $n+1$ and given by:
\begin{equation}
	B_n(x) =  e^{- n m x} \, , e^{- (n-2)m x} \,, \ldots \, , e^{ i(n-2) mx } \,, e^{i n mx}
\end{equation}
This set of functions is the space of solutions to the equation:
\begin{equation}
	\begin{split}
	  \p_x^{\frac{1+(-1)^n}{2}}\prod_{k=0}^{ \lfloor n/2 \rfloor}\Big(\hat\Box +(n-2k)^2 m^2\Big) \prod_{\pm}^n G_{\pm m} = 0
	\end{split}
\end{equation}
The $\p_x$ factor is responsible for the constant solutions which appear for even $n$.
\\

As we can see the order drops greatly - from an exponential in $n$ to linear. This happens through factorization of the unequal mass operator. For example $n=2$:
\begin{equation}
	\Box^4 \left( \Box+4m^2 \right)^4 =  \Box^3 (\Box+4m^2)^3 \partial_x \left( \partial_x \left(\Box+4m^2\right) \right)
\end{equation}
 When $2\leq k \leq n-1 $ of $n$ masses are the same, the order is $2^{n-k-1}(k+1)$, when there are clusters of $k_s$ equal masses, $\sum_s k_s=n $,
then the order in $\p_x$ is $\prod_s (k_s+1)$, modulo
possible coincidences at $\sum_i \pm m_i=0$ for several even $k_s$.

\subsection{Beyond $D=1$}

Beyond $D=1$ explicit dependence on $\vec x^2$ appears
and equations become far more complicated. The most practical way to write them is to use the dilatation operator:
\begin{equation}
		\hat{\Lambda} =  x^\mu \p_{x^\mu}
\end{equation}
together with the Laplacian
\begin{equation}
\hat \Box:=\p_\mu \p^\mu
\end{equation}
Then the procedure to obtain the {\bf equation on a product of solutions}
of the original $(\Box +m^2)G=0$, outlined in \cite{Mishnyakov:2023wpd},
gets a straightforward and easily computerizable realization.
By definition, the order of the resulting equation in $\hat\Lambda$ remains the same: from $n+1$ for equal masses and
$2^n$ for all masses different.
\\
For example for $n=2$ and equal masses one has:
\begin{equation}\label{eq:n2equalmassposition}
	 \hat{\mathcal{E}_x^{(2)}}= \Lambda (\Box+4m^2) + (D-1)\left(\Box + 2m^2\right)
\end{equation}

As mentioned above, ar $D\neq 1$ the equation is also graded by powers of $x=\sqrt{x^\mu x_\mu}$. For further purposes of studying the Fourier transform, it is important to keep track of this grading as well. For the equal mass case it was studied in \cite{Mishnyakov:2023wpd} - the order is $n-1$ if we also count the power of $x$ that enters the operator $\Lambda$. The same question for the unequal mass case still remains unresolved.
\\

Note that for the action on invariant functions of $x=\sqrt{x^\mu x_\mu}$  the operators $\Lambda$ and $\Box$ can be rewritten as:
\begin{equation}
	\Lambda = x \dfrac{\partial}{\partial x } \, , \quad \Box  = \dfrac{\partial^2}{\partial x^2} + \dfrac{D-1}{x} \dfrac{\partial}{\partial x}
\end{equation}
where the derivative is also over $x=\sqrt{x^\mu x_\mu}$. We will use this representation to count the order of the equations in $x$ and derivatives. That way it is also easier to compare with $D=1$. For other purposes the representation in terms of $\Lambda, \Box$ and $x^2$ seems to be more appropriate. This hierarchy of orders of the equations is summarized in the following table (see Appendix 2 of \cite{Mishnyakov:2023wpd}), where we list the leading term in both $x$ and $p_x$:
\vspace{0.3cm}
\begin{center}
	\begingroup
	\centering
	\renewcommand{\arraystretch}{2.5}
	\newcolumntype{M}[1]{>{\centering\arraybackslash}m{#1}}
\begin{tikzpicture}
	\node (table) [inner sep=0pt] {
		\begin{tabular}{M{2cm}|M{4cm}|M{4cm}}
			
			\boxed{\hat{\cal E}_x}   & \text{all masses equal} &   \text{all masses different}  \\
			\hline
			$D=1$ &    $\p_x^{n+1}$ &  $\p_x^{2^n}$ \\
			\hline
			$ D >1$ &   $x^{n-1}\p_x^{n+1}$ & $x^2 \p_x^4 \ {\rm at}\ n=2$ 	\vspace{0.15cm}	 \newline
			\vspace{0.15cm}	 $x^{10} \p_x^8 \ {\rm at}\ n=3$
			\newline
			\vspace{0.15cm}	 $x^{48} \p_x^{16} \ {\rm at}\ n=4$
			\newline
			\vspace{0.15cm}	 $x^{226} \p_x^{32} \ {\rm at}\ n=5$
			\newline	 $x^{?} \p_x^{2^n} $ \\
			
		\end{tabular}
	};
	\draw [rounded corners=.3em,line width=0.4mm] (table.north west) rectangle (table.south east);
\end{tikzpicture}
\endgroup
\end{center}
\vspace{0.3cm}

The simplest way to derive these equations is to act by powers of $\hat\Lambda$ on the product
of Green functions $Z_n = \prod_{i=1}^n G_{m_i}(x^2)$.
Then, making use of the Leibnitz rule and the equation $\hat\Lambda^2 G_m = -(D-2)\hat\Lambda G_m - m^2x^2 G_m$
for individual Green functions,
we obtain the relations of the form
\be
\hat\Lambda^s Z_n = \sum_{\nu_1,\ldots, \nu_n=0,1} a_s^{\{\nu_1,\ldots, \nu_n\}}
\overbrace{\prod_{i=1}^n \hat\Lambda^{\nu_i} G_{m_i}}^{Z_{\nu_1,\ldots, \nu_n}}
\label{LsZn}
\ee
Since there are $2^n$ in dependent quantities $Z_{\nu_1,\ldots, \nu_n}$ we conclude that the
determinant of the matrix of the size $2^n+1$ should vanish, e.g.

{\footnotesize
	\be
	\hat \Lambda^2 G = a_2^{\{1\}} \hat \Lambda G + a_2^{\{0\}} G
	\ \ \Longleftrightarrow \ \
	\left(\begin{array}{ccc} 0 & 1 & -Z_1 \\ 1 & 0 &-\hat \Lambda Z_1 \\ a_2^{\{1\}} & a_2^{\{0\}} & -\hat\Lambda^2 Z_1 \end{array}\right)
	\left(\begin{array}{c} \hat\Lambda G \\ G \\ 1 \end{array}\right) = 0
	\ \ \Longleftrightarrow \ \
	{\rm det} \left(\begin{array}{ccc} 0 & 1 & -Z_1 \\ 1 & 0 &-\hat \Lambda Z_1 \\ a_2^{\{1\}} & a_2^{\{0\}}
		& -\hat\Lambda^2 Z_1 \end{array}\right)
	= 0
	\nn
	\ee
}
\vspace{0.2cm}

\noindent
for the zero-loop case $n=1$ with $Z_1=G$.
This provides the equation in coordinate space
in the form of linear relation between $\hat\Lambda^s Z_n$ with $s=0,\ldots, 2^n$,
which can be easily Fourier-transformed to the momentum space.
It is clearly of degree $2^n$ in $\p_x$, but the degree in $x^2$ is a little trickier.
In the first approximation we can use the fact that the coefficients $a_s$ in (\ref{LsZn})
are of degree $\left[\frac{s}{2}\right]$ in $x^2$, what implies that the degree of determinant
cannot exceed than $\sum_{s=0}^{2^n} \left[\frac{s}{2}\right] =  4^{n-1}$.
However, there are two corrections: on the one hand, the actual power is {\it slightly} lower because of the overlaps
between the powers of $x^2$ in different lines, and on the other hand the entire determinant appears to be proportional
to some power of $x^2$, which can be factored out -- actually it is approximately one half of the net $4^n$.
Moreover, the coefficient of $\hat\Lambda^0 Z_n$ also appears to be proportional to $x^2$,
and, since the action of $\hat\Lambda$ on functions of $x^2$ is always proportional to $x^2$,
$\hat\Lambda F(x^2) = 2x^2F'(x^2)$, this allows  to subtract one more power.
This is a small correction, but it is essential for exact answers in above table.
The resulting power of $x$ is {\it slightly} lower than $4^{n-1}$, the general answer is still unavailable
(and even the power $226\approx 4^4=256$ is a result of a hard calculation).
\be
\begin{array}{c|c|c|c|c}
	n &  4^{n-1} & x^2-\text{power of determinant} & \text{common factor} &x^2-\text{power of the equation}    \\
	\hline
	1 & 1 &1 &  0 & 1-1=0 \\
	\hline
	2 & 4 & 4 &  1 & 2-1=1 \\
	\hline
	3 & 16 & 12 & 6 & 6-1=5\\
	\hline
	4 & 64 & 52 & 27 & 25-1=24 \\
	\hline
	5 & 256 & ? & ? & 114-1=113 \\
	\hline
	\cdots &&&&
\end{array}
\nn
\ee

For explicit examples of coordinate-space equations for banana diagrams
see the appendices in \cite{Mishnyakov:2023wpd}.

\subsection{Transformation of operators to momentum space}

In order to go to momentum space we can do the Fourier transform directly in terms of the operators. The rules follow from considering functions of invariants $\vec x^2$ and $t=\vec p^2$:
\begin{equation}
	\begin{split}
			\dfrac{\partial}{\partial x^\mu} \int e^{i p_\mu x^\mu} f(t) &= \int e^{i p_\mu x^\mu} \left( i p_\mu \right)f(t)
			\\
			 x^\mu  \int e^{i p_\mu x^\mu} f(t) &= \int e^{i p_\mu x^\mu}  \left( i 	\dfrac{\partial}{\partial p_\mu}\right) f(t)
	\end{split}
\end{equation}
Hence, for invariant operators one has the following list  of rules for transiting to momentum space:
\vspace{0.3cm}
\begin{center}
	\begingroup
	\centering
	\renewcommand{\arraystretch}{2.5}
	\newcolumntype{M}[1]{>{\centering\arraybackslash}m{#1}}
	\begin{tabular}{M{4cm}|M{4cm}}
		
		 \text{Coordinate space} &   \text{Momentum space}  \\
		 \hline
			$\Box = \partial_\mu \partial^\mu $ &  $ -t = - p_\mu p^\mu $ \\
		$\Lambda = x\partial_x = x^\mu \partial_\mu $ &  $ - (D-2) -2  \dfrac{\partial}{\partial t}t$ \\
	
		$-(D+\Lambda)$ &  $ 2 t \dfrac{\partial}{\partial t} = p^\mu \dfrac{\partial}{\partial p^\mu}$ \\

		   $x^2 = x^\mu x_\mu$ & $-2D \dfrac{\partial}{\partial t}-4 t \dfrac{\partial^2}{\partial t^2} $
	\end{tabular}
	\endgroup
\end{center}
\vspace{0.3cm}

This allows us to compute the operators $\hat{E}_t^{(n)}$. These operators now define the differential equations for the momentum space banana integral, which is now indeed a loop integral and not a product of propagators. One can explicitly show at low loop orders, that the result of integration indeed satisfies the equation. Some examples were considered in \cite{Mishnyakov:2023wpd}, and some more will appear in sec. \ref{sec:Factor}. Here we limit ourselves again to the $n=2$ equal mass case:
\begin{equation}\label{eq:n2equalmassmomentum}
		\hat{E}_t^{(2)}=\left(D+2t\frac{d}{dt}\right)(t-4m^2) -2(D-1)(t-2m^2)=
		\left( 2t(t-4m^2)\dfrac{d }{dt}+(t(4-D)-4m^2) \right)
.\end{equation}
 which should be compared to \eqref{eq:n2equalmassposition}.
 \\

Here we do not intend to study the solutions of the equations; instead, as for the momentum space we want to study their order for different cases. It is clear from the table above, that each multiple of $x$, whether it enters through $x^2$ or as part of the $\Lambda$ operator, contributes a $t$ derivative. Position space derivatives on the other hand correspond just to powers of $t$. Having that in mind we can  represent the order of the resulting equation in the following table:

\begin{center}
	\begingroup
	\centering
	\renewcommand{\arraystretch}{2.5}
	\newcolumntype{M}[1]{>{\centering\arraybackslash}m{#1}}
	\begin{tikzpicture}
		\node (table) [inner sep=0pt] {
			\begin{tabular}{M{2cm}|M{4cm}|M{4cm}}
				
				\boxed{\hat{E}_t}   & \text{all masses equal} &   \text{all masses different}  \\
				\hline
				$D=1$ &    $p^{n+1}$ &  $p^{2^n}$ \\
				\hline
				$ D >1$ &   $t^{n} \dfrac{\partial^{n-1}}{\partial t^{n-1}}$ & $t^2 \dfrac{\partial^{2}}{\partial t^{2}}\ {\rm at}\ n=2$ 	\vspace{0.15cm}	 \newline
				\vspace{0.15cm}	 $t^4 \dfrac{\partial^{10}}{\partial t^{10}} \ {\rm at}\ n=3$
				\newline
				\vspace{0.15cm}	 $t^8 \dfrac{\partial^{48}}{\partial t^{48}} \ {\rm at}\ n=4$
				\newline
				\vspace{0.15cm}	 $t^{16} \dfrac{\partial^{226}}{\partial t^{226}} \ {\rm at}\ n=5$
				\newline	 $t^{2^{n-1}} \partial_t^{?}  $ \\				
			\end{tabular}
		};
		\draw [rounded corners=.3em,line width=0.4mm] (table.north west) rectangle (table.south east);
	\end{tikzpicture}
	\endgroup
\end{center}
\vspace{0.3cm}
An interesting observation can be made about the order of the equations. For the equal mass case the order of the differential operator is $n-1$, whereas the position space operator had order $n+1$. On the other hand, we expect that the solutions to the momentum space equations are just the Fourier transforms of the momentum space equations, so we expect the number of solutions to be the same. In fact, the counting appears to be more intricate. This can already be seen for $D=1$. At $D=1$ the equations are not differential, just as for the single cut propagator we have:
\begin{equation}
	(p^2-m^2) G(t) = 0
\end{equation}
On the other hand we have two solutions to the equations of motion which are clearly given by:
\begin{equation}
	G(p) =\delta(p \pm m)
\end{equation}
We will not go further into this question here and leave it for a more technical paper. The takeaway, however, is that comparing the orders of $\hat{\mathcal{E}_x}$ and $\hat{E}_t$ is not naive.
\\

The unresolved piece of the puzzle is the lower right corner of the table. The degree of the differential operator $\hat{E}_t$ for unequal masses is unknown as it corresponds to the power of $x$ in the respective position space equation. The correspondence between the two unequal and equal mass case is again that of factorization.

\section{PF equations in the momentum space}

Our approach allows to derive differential equations for momentum space integrals starting only with the equations of motion. Another commonly discussed approach is to derive Picard-Fuchs equations for the parametric representation of the integral, which will produce some differential operator in the $t$ variable $\widehat{PF}$. In this section we will discuss the origin and form of $\widehat{PF}$ and discuss it's properties. Most of the results obtained in this section can also be found in \cite{Lairez:2022zkj, Adams:2015pya}

\subsection{The origin and  the form of PF equations}
The parametric representation for momentum Feynman integrals and banana integrals specifically is well known and relies on the 2 identities:
\begin{equation}
	\begin{split}
		\dfrac{1}{p^2-m^2} = \int_{0}^{\infty} d\alpha\, e^{-\alpha (p^2-m^2)}
		\\
		\delta(p^2-m^2) = \int_{-\infty}^{\infty} d\alpha \, e^{-i \alpha (p^2-m^2)}
	\end{split}
\end{equation}
For general derivations see \cite{Weinzierl:2022eaz}. The momentum space banana diagram is an $n-1$ loop integral and its parametric form is given by:
\be
I_n(t):= \int   \delta\left(\sum_{i=1}^n\vec k_i - \vec p\right)\prod_{i=1}^n \delta(k_i^2-m_i)^2 d^D\vec k_i
= \int\limits_{\mathbb{R}^n} \frac{ U(\alpha)^{\frac{n(2-D)}{2}}\delta\left(\sum_{i=1}^n\alpha_i-1\right)\prod_{i=1}^{n} d\alpha_i    }
{ \Big(\alpha_1\ldots \alpha_n \cdot t - U(\alpha)\cdot\sum_{i=1}^n \alpha_im_i^2\Big)^{1+\frac{(n-1)(2-D)}{2}}  }
\label{FIn}
\ee
where
\be
U(\alpha) := \prod_{i=1}^n \alpha_i \sum_i \alpha_i^{-1} = \sum_{i=1}^n \alpha_1\ldots\check \alpha_i \ldots \alpha_n
\ee
and the denominator is usually denoted as:
\begin{equation}
	F(\alpha) = \Big(\alpha_1\ldots \alpha_n \cdot t - U(\alpha)\cdot\sum_{i=1}^n \alpha_im_i^2\Big)
\end{equation}
The $U$ and $F$ are called the first and second Symanzik polynomials correspondingly. The integration goes over the whole $\mathbb{R}^n$ (i.e. from $-\infty$ to $+\infty$) as opposed to only $\alpha_i>0$ since we are considering the maximal cut.
\\
After complexification the integral can be written as a contour integral:
\begin{equation}\label{eq:cutcontour}
	I_n(t)= \oint\limits_{\gamma} \prod_{i=1}^{n-1} d\alpha_i  \left.  \dfrac{ U(\alpha)^{\frac{n(2-D)}{2}}   }{ \Big(\alpha_1\ldots \alpha_n \cdot t - U(\alpha)\cdot\sum_{i=1}^n \alpha_im_i^2\Big)^{1+\frac{(n-1)(2-D)}{2} }  } \right|_{\alpha_n = 1 -\sum_{i=1}^{n-1} \alpha_i}
\end{equation}
where $\gamma$ is some closed contour appropriately encircling the zero set of $F$. We could also work projectively to get:
\begin{equation}
	I_n(t)=\int\limits_{\Gamma_n} \frac{ U(\alpha)^{\frac{n(2-D)}{2}}\omega }
	{ \Big(\alpha_1\ldots \alpha_n \cdot t - U(\alpha)\cdot\sum_{i=1}^n \alpha_im_i^2\Big)^{1+\frac{(n-1)(2-D)}{2}}  }
	\label{ProjIn}
\end{equation}
where $\omega=\sum_{i=1}^{n}(-1)^{i+1}\alpha_i d\alpha_1\wedge\dots\wedge\widehat{d\alpha_i}\wedge\dots\wedge d\alpha_n$ is measure on $\mathbb{CP}^{n-1}$ and $\Gamma_n=\left\{(\alpha_1,\dots,\alpha_n)\in\mathbb{CP}^{n-1}\vert \alpha_i\in \mathbb{R}\right\}$. It is important to note, that the integrand in \eqref{ProjIn} equals the one in \eqref{FIn} only on the integration domain. The contour $\Gamma_n$ is closed and to simplify calculations we can rewrite this integral as an integral over $\mathbb{C}^n$:
\begin{equation}
	I_n(t)=\oint\limits_{\bar{\gamma}} \frac{ U(\alpha)^{\frac{n(2-D)}{2}}\prod_{i=1}^n d\alpha_i }
	{ \Big(\alpha_1\ldots \alpha_n \cdot t - U(\alpha)\cdot\sum_{i=1}^n \alpha_im_i^2\Big)^{1+\frac{(n-1)(2-D)}{2}}  }
	\label{ComplexIn_gamma}
\end{equation}
where $\bar{\gamma}$ is the fibration over $\gamma$ with fibre $S^1$. This equation instantly follows from the identity $(\frac{d\alpha_1}{\alpha_1}+\dots+\frac{d\alpha_n}{\alpha_n})\wedge\omega=\prod_{i=1}^n d\alpha_i$ and integration over the fibres.
Thus in any case the integral is a ``period'' and satisfies the Picard-Fuchs equation.
Denoting the integrand $n$-form by $\Omega$, we can say that the equation:
\be
\widehat{PF} \int\limits_{\bar{\gamma}} \Omega = 0 \ \ \Longleftarrow \ \ \widehat{PF} \cdot \Omega = d\beta
\label{PForigin}
\ee
follows from a differential equation where the l.h.s. contains derivatives in $t$,
while the r.h.s. is made from $\alpha$-derivatives.\\

\subsection{The case of equal masses}

Notice that for $D=2$ the parametric integral becomes simpler:
\begin{equation}
	I_n(t) = \oint_\gamma \prod_{i=1}^{n-1} d\alpha_i \dfrac{1}{F(\alpha)}
\end{equation}
In this case it is straightforwardly understood as a period of the manifold that is the vanishing locus of the denominator:
\begin{equation}
	F(\alpha) = 0
\end{equation}
It turns out to be a singular (for $n>3$) $n-2$ dimensional Calabi-Yau variety. As for our matter of deriving a differential equation this case is also simpler. For example, at $n=2$ (one-loop) we have:
\begin{equation}\label{eq:oneloopeqmass}
	\left\{t\,(t-4m^2)\frac{\p}{\p t}
	+ \Big(\underline{ t-2m^2}\Big)\right\} \left.\dfrac{1}{F}\right|_{\alpha_2=1-\alpha_1} = m^2 \left(  \dfrac{\partial}{\partial \alpha_1} \right) \left. \dfrac{1-2 \alpha_1}{F(\alpha_1,\alpha_2)} \right|_{\alpha_2=1-\alpha_1}
\end{equation}
For $D=2$, the underlined coefficient  $t-2m^2$  appears equal to
one half of the $t$-derivative of the first coefficient $t(t-4m^2)$.
Moreover, this property is true for bigger number of loops,
where the coefficients of the PF equation for $D=2$ and all masses equal are given in table \ref{table:eqmassD2}.
\vspace{0.1cm}
\begin{table}[h]
		\vspace{0.3cm}
\begingroup
\centering
\renewcommand{\arraystretch}{2.5}
\newcolumntype{M}[1]{>{\centering\arraybackslash $}m{#1}<{$}}
\footnotesize
	\begin{tabular}{M{0.6cm}|M{0.3cm}|*{4}{M{3cm}|} M{0.6cm}}
		n-1 & \ldots &\frac{\p^4}{\p t^4}&\frac{\p^3}{\p t^3}&\frac{\p^2}{\p t^2}&\frac{\p}{\p t} & 1  \\
			\hline
		1& &&&&
		t(t-4) &\underline{ t-2} \\ \cline{1-1}\cdashline{2-2}\cline{3-7}
		2 &&&& t(t-1)(t-9)=
		\newline 
		=t^3-10t^2+9t& \underline{3t^2-20t+9} & t-3 \\ \cline{1-1}\cdashline{2-2}\cline{3-7}
		3&&&  t^2(t-4)(t-16)= \newline =t^4-20t^3+64t^2 & \underline{6t^3-90t^2+192 t}   & 7t^2-68t+64 & t-4  \\ \cline{1-1}\cdashline{2-2}\cline{3-7}
		4&&  \begin{minipage}{3cm}
			$t^2(t-1)(t-9)(t-25)$ \newline \centering
				$\rotatebox{90}{=}$
			 \newline $t^5-35t^4+259t^3-225t^2$
		\end{minipage}
		& \underline{2\cdot (5t^4-140t^3+}\newline \underline{+777t^2-450t)} & 25t^3-518t^2+1839t-450 & 25t^2-196t+285  & t-5\\ \cline{1-1}\cdashline{2-2}\cline{3-7}
		\vdots &\vdots &\vdots &\vdots &\vdots & \vdots & \vdots
	\end{tabular}

\endgroup
\caption{Coefficient of the equal mass banana Picard-Fuchs operators. The underlined coefficient is always two times the derivative of the first coefficient $c_{n-2}=\frac{n-1}{2}\frac{\partial c_{n-1}}{\partial t}$, where $n-1$ is the number of loops. We have set $m=1$ for brevity }
		\label{table:eqmassD2}
		\vspace{0.3cm}
\end{table}
The general structure of the operator is:
\begin{equation}
	D=2 \ : \ \widehat{PF}^{(2)} =  c_n \dfrac{\partial^{n-1}}{\partial t^{n-1}} + \sum_{i=0}^{n-2} c_i \dfrac{\partial^{i}}{\partial t^i}
\end{equation}
This agrees with the those from \cite{Lairez:2022zkj} and can be compared to momentum space operators in \cite{Mishnyakov:2023wpd}. On the level of PF equations the generalization to $D \neq 2$ is straightforward and $D$ appears polynomially in the coefficient of the PF operator. Once again, for $n=2$  and all masses equal:
\begin{equation}
	\left\{t\,(t-4m^2)\frac{\p}{\p t}
	+ \Big(\underline{(N+1) t-2m^2}\Big)\right\} \left.\frac{U^{2N}}{F(\alpha_1,\alpha_2)^{N+1}} \right|_{\alpha_2=1-\alpha_1}
	= m^2\left(\frac{\p}{\p \alpha_1}\right) \left.\frac{1-2\alpha_1}{F(\alpha_1,\alpha_2)^{N+1}} \right|_{\alpha_2=1-\alpha_1}
\label{PDn2eqmD}
\end{equation}
with $N = \frac{2-D}{2}$. Hence:
\begin{equation}
	\widehat{PF}_2 = t\,(t-4m^2)\frac{\p}{\p t}
	+ \left( \dfrac{4-D}{2} t-2m^2 \right)
\end{equation}

It is worth mentioning at this point already that for equal masses but generic $D$ the operators $\hat{E}^{(n)}_t$ are equal to $\widehat{PF}^{(n)}$. This statement was confirmed on all the calculated examples. Moreover at $D=2$ \cite{Mishnyakov:2023wpd} contains a comparison of the leading and subleasing coefficient with the generic formulas of \cite{Lairez:2022zkj}. Hence the Fourier transform is an easier method to compute these operators for many orders, when they are needed explicitly for analysis. We have the first 8 loops of the position space operators listed in \cite{Mishnyakov:2023wpd}

\subsection{Different masses}
Again, the simplest case is $D=2$. One can explicitly obtain the PF operators and see that they are of the form
\begin{equation}
	D=2: \ \ \
	\widehat{PF}^{(n)} =\underbrace{ \tilde{c}_{\operatorname{ord}(n)-1}\prod_{\pm} \Big(t-(m_1\pm m_2 \pm\ldots \pm m_n)^2\Big)  }_{c_{\operatorname{ord}(n)-1}}
	\frac{\p^{\operatorname{ord}(n)-1}}{\p t^{\operatorname{ord}(n)-1}}
	+ \sum_{i=1}^{\operatorname{ord}(n)-1} c_{i} \frac{\p^{i} }{\p t^{i}}
	\label{PFdifmD2}
\end{equation}
The structure of other coefficients is much more complicated now, however, what is important is that the order also changes compared to the equal mass case. In particular it was conjectured in \cite{Lairez:2022zkj} that the order of the PF equation at $D=2$ for generic masses is:
\begin{equation}
	\operatorname{ord}(n) = 2^n- \binom{n+1}{\lfloor \frac{n+1}{2} \rfloor}
\end{equation}
This leads to exponentially growing order, which is closer to agreement with how drastically the order grows for the $\hat{E}_t^{(n)}$ operators.
\\

Interestingly enough, the generalization to $D \neq 2$ is now non-trivial. The only example, for which $D$ only deforms in the coefficients, is for $n=2$:
\begin{equation*}
	\widehat{PF}^{(2)} = 	\left(2t(t-(m_1-m_2)^2)(t-(m_1+m_2)^2) \frac{\partial}{\partial t}
	+ \left( (4-D)t^2-2(m_1^2+m_2^2)t+ (D-2)(m_1^2-m_2^2)^2 \right) \right)
\end{equation*}

\be
	\left\{2t(t-(m_1-m_2)^2)(t-(m_1+m_2)^2) \frac{\partial}{\partial t}
 + \left( (4-D)t^2-2(m_1^2+m_2^2)t+ (D-2)(m_1^2-m_2^2)^2 \right)\right\}
\cdot \nn
\ee
\vspace{-0.5cm}
\be
\!\!
\cdot
\frac{(\alpha_1+\alpha_2)^{2-D}}
{\Big((\alpha_1+\alpha_2)(m_1^2\alpha_1+m_2^2\alpha_2)-t\alpha_1\alpha_2 \Big)^{2-\frac{D}{2}}}
= \nn \\
=2\left(m_2^2(t+m_1^2-m_2^2)\frac{\partial}{\partial \alpha_1}+m_1^2(t+m_2^2-m_1^2)\frac{\partial}{\partial \alpha_2}\right)\frac{(\alpha_1+\alpha_2)^{3-D}}
{\Big((\alpha_1+\alpha_2)(m_1^2\alpha_1+m_2^2\alpha_2)-t\alpha_1\alpha_2\Big)^{2-\frac{D}{2}}}
=\nn \\
=\frac{\p}{\p \alpha_1}
\frac{2\Big(-m_1^2(t-m_1^2+m_2^2)\alpha_1+m_2^2(t+m_1^2-m_2^2)\alpha_2\Big)(\alpha_1+\alpha_2)^{2-D}}
{\Big(t\alpha_1\alpha_2-(\alpha_1+\alpha_2)^{2-D} (m_1^2\alpha_1+m_2^2\alpha_2) \Big)^{2-\frac{D}{2}}}
\ee
where $\alpha_2=1-\alpha_1$.
In fact, after this substitution the expression is simplified a little, but we keep it in this
form, which is  generalizable to higher loops.
The l.h.s. of this one-loop equation is in accordance with \cite{Muller-Stach:2012tgj}.
\\

At higher $n$ going to $D\neq 2$ results in the growth of the order of the equation. In particular for $n=3$ the equation is of order 4:
\begin{equation}
	D\neq 2 \, : \quad   \widehat{PF}^{(3)}	 = \sum_{i=0}^{4} c_i(m_1,m_2,m_3,D) \dfrac{\partial^i}{\partial t^i}
\end{equation}
This is in agreement with \cite{Adams:2015pya} and we can independently confirm the validity of their answer. To the best of our knowledge there is no prediction on the order of the PF equation at $D \neq 2$ and general $n$. Note that for the cohomological reasoning about the closed contour integral to be valid we should require that $D$ is even. On the other hand after computation we clearly see that  dependence of the operator coefficients on $D$ is analytic.  Finally we summarize what is known about Picard-Fuchs equations for banana-graphs in the table:

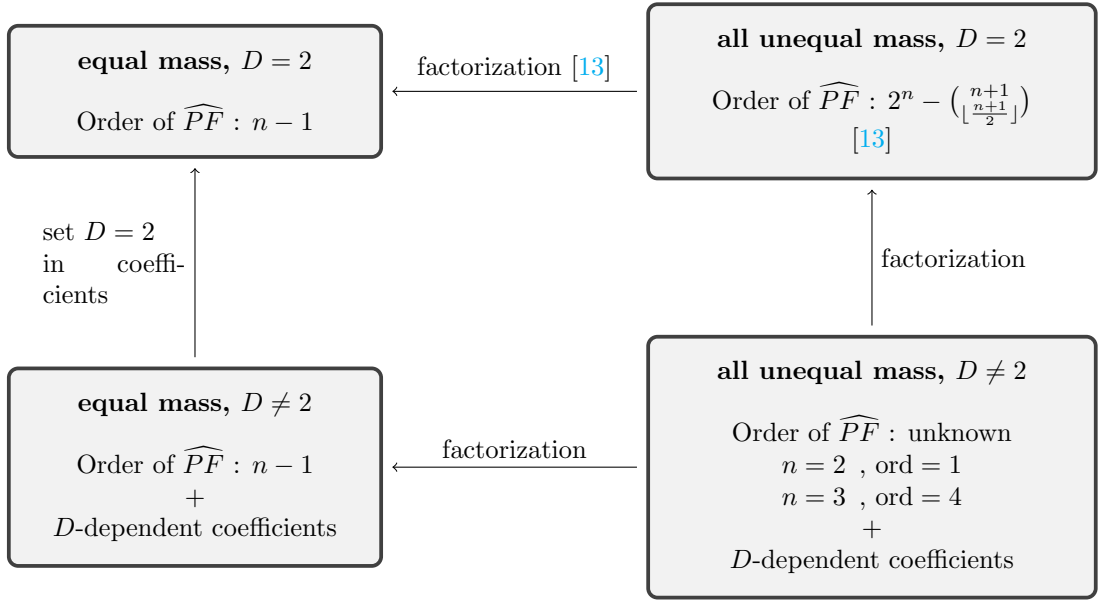
\vspace{0.3cm}
\begin{center}

	\begin{tikzpicture}
		\node (A) at (0,0){\begin{minipage}{5cm}
				\begin{tcolorbox}
					\centering
					{\bf equal mass, $D=2$}
					\\
					\vspace{0.4cm}
					Order of $\widehat{PF}$  : $n-1$
				\end{tcolorbox}
			\end{minipage}
	};
		\node (B) at (9,0){
			\begin{minipage}{6cm}
				\begin{tcolorbox}
						\centering
					{\bf all unequal mass, $D=2$}
					\\
					\vspace{0.4cm}
					Order of $\widehat{PF}$  : $ 2^n- \binom{n+1}{\lfloor \frac{n+1}{2} \rfloor}$ \, \cite{Lairez:2022zkj}
				\end{tcolorbox}
			\end{minipage}
	};
	\node (C) at (0,-5){
	\begin{minipage}{5cm}
		\begin{tcolorbox}
				\centering
			{\bf equal mass, $D \neq 2$}
			\\
			\vspace{0.4cm}
			Order of $\widehat{PF}$  : $n-1$
			\\
			+
			\\
			$D$-dependent coefficients
		\end{tcolorbox}
	\end{minipage}
};
\node (D) at (9,-5){
	\begin{minipage}{6cm}
		\begin{tcolorbox}
					\centering
			{\bf all unequal mass, $D \neq 2$}
			\\
			\vspace{0.4cm}
			Order of $\widehat{PF}$  :  unknown
			\\
			$n=2$ \,, $\operatorname{ord}=1$
			\\
			$n=3$ \,, $\operatorname{ord}=4$
			\\
			+
			\\
			$D$-dependent coefficients
		\end{tcolorbox}
	\end{minipage}
};
\draw[->] (B) -- node[midway,above] {factorization \cite{Lairez:2022zkj}} (A) ;
\draw[->] (C) -- node[midway,left] { \begin{minipage}{1.9cm}
		set $D=2$\\ in coefficients
	\end{minipage} } (A) ;
\draw[->] (D) -- node[midway,above] { factorization } (C) ;
\draw[->] (D) -- node[midway,right] { factorization } (B) ;
	\end{tikzpicture}
\end{center}
\vspace{0.3cm}

The most complicated equation is the bottom right square. Its reduction to the equal mass or $D=2$ cases is archived by factorization. However even looking at the orders we see that these are two separate phenomena.

\subsection{Self-adjoint property}
At $D=2$ the Picard-Fuchs operators correspond to periods of Calabi-Yau manifolds and hence are supposed to be self-adjoint \cite{bogner2013algebraic,almkvist2010art,almkvist2013class,van2017calabi}. Generically, for operator:
\begin{equation}
	P=\sum_i a_i\frac{\partial^i}{\partial t^i}
\end{equation}
Its adjoint is
\begin{equation}
	\overline{P}=\sum_i (-1)^i \frac{\partial^i}{\partial t^i}a_i
\end{equation}
Therefore the operator is self-adjoint when the following relation holds:
\begin{equation}
	\label{adjoint_eq}
	 P g(t)=(-1)^{\text{deg} P}\overline{P g(t)}=(-1)^{\text{deg} P}g(t)\overline{P}
\end{equation}
for some $g(t)\neq 0$. For $g(t)=1$ this implies special explicit relations between all coefficients with even $i$, e.g.
\begin{equation}
	\begin{split}
			a_{n-4} =& \frac{(n-1)(n-2)(n-3)}{12} \cdot \dfrac{\partial^3 a_{n-1}}{\partial t^3}
		- \frac{(n-2)(n-3)}{4}\cdot \dfrac{\partial^2 a_{n-2}}{\partial t^2} + \frac{n-3}{2} \cdot \dfrac{\partial a_{n-3}}{\partial t}
		=
		\\
		&= - \frac{(n-1)(n-2)(n-3)}{24} \cdot \dfrac{\partial^3 a_{n-1}}{\partial t^3}  + \frac{n-3}{2} \cdot \dfrac{\partial a_{n-3}}{\partial t}
	\end{split}
\end{equation}
The PF operators are self-adjoint for $D=2$ and generic masses, which is in accordance with the expectation that the Feynman integrals are periods of $(n-2)$-dimensional Calabi-Yau variety. In particular for equal mass $D=2$ operators (see table \ref{table:eqmassD2}):
\begin{equation}\label{eq:selfadjointforequalmassD2}
	c_{n-2} = \dfrac{n-1}{2}\,  \dfrac{\partial c_{n-1}}{\partial t}
\end{equation}
\\

For generic dimensions this property is broken. This implies that PF operators for generic banana diagrams are {\it not} self-adjoint and are not of the Calabi-Yau type. This is clear already from the equal mass case for $D \neq 2$. For a nontrivial example, we can look at $n=4$:
\begin{equation}\label{eq:PF4}
	\begin{split}
		\widehat{PF}^{(4)}&=t^2 \left(t-16 m^2\right) \left(t-4m^2\right)\dfrac{\p^3}{\p t^3} -3  t \left(-10 (D-5)	m^2 t+(D-4) t^2-64 m^4\right)\dfrac{\p^2}{\p t^2} +
		\\
		&+ \left(-16 (D-4) D m^4+((88-7 D) D-216) m^2
		t+\frac{1}{4} (D-4) (11 D-36) t^2\right)\dfrac{\p}{\p t}  - 
		\\
		& -\frac{1}{4} (D-3) (3 D-8) \left(2 (D+2) m^2+(D-4) t\right)
	\end{split}
\end{equation}
which can be checked to be not self-adjoint. Moreover, according to \cite{Mishnyakov:2023wpd} (see also next section the deformation $D\neq 2$ of \eqref{eq:selfadjointforequalmassD2} is:
\begin{equation}\label{eq:coefrelforequallmassDnot2}
	c_{n-2}= \dfrac{(n-1)}{2} \cdot \left( \dfrac{n(D-2)}{2t} c_{n-1} -(D-3) \dfrac{\partial c_{n-1} }{\partial t} \right)
\end{equation}
which true for all loops. The divergence of \eqref{eq:coefrelforequallmassDnot2} from \eqref{eq:selfadjointforequalmassD2} on itself does not yet mean that the operator is not self adjoint. We could potentially find a non-trivial $g$ from \eqref{adjoint_eq}. However, if a simple check shows that system of equations on all the coefficients is inconsistent, hence the operator is not self-adjoint.

 On the other hand the rather simple structure \eqref{eq:coefrelforequallmassDnot2} suggest there could exist an appropriate deformation of the self-duality condition. We leave this for future investigation.
\\

\section{Relation between Fourier transform of position space equations and Picard-Fuchs equations} \label{sec:Factor}

As clear from the tables above the operators $\hat{E}^{(n)}_t$ and $\widehat{PF}^{(n)}$ are significantly different from the equal mass case. One can clearly see, that for unequal masses the order of the operator obtained by Frourier transfrom of the position space equations grows faster than those of the Picard-Fuchs operator. On the other hand one can check that the kernel of the PF operator is contained in the kernel of $\hat{E}^{(n)}_t$. This means that the operators are once again related by factorization. Why and how this happens is yet unclear and remains a problem for future investigation.
\\

The only truly tractable example is at $n=2$. In position space one has \cite{Mishnyakov:2023wpd}:
\begin{equation}
	\hat{\mathcal{E}}^{(2)}_x = x^2 \Big(\Box+(m_1+m_2)^2\Big)\Big(\Box + (m_1-m_2)^2\Big)
	+ 2(D-1)\hat\Lambda(\Box+m_1^2+m_2^2)
	+ 2(D-1)(D-2)(\Box+m_1^2+m_2^2)
\end{equation}
and it's momentum space version
\begin{equation}
	\begin{split}
		&\hspace{5cm}\hat{E}_t^{(2)} = -2\frac{\p}{\p t} \left( \widehat{PF}^{(2)}\right) =
		\\
		& -2\frac{\p}{\p t} 	\left(2t(t-(m_1-m_2)^2)(t-(m_1+m_2)^2) \frac{\partial}{\partial t}
		+ \left( (4-D)t^2-2(m_1^2+m_2^2)t+ (D-2)(m_1^2-m_2^2)^2 \right) \right)
	\end{split}
\end{equation}
Here we can clearly trace the reduction of the order of operators from $2$ to $1$ according to the tables. Even at $n=3$ we seem to lack the appropriate language to describe the relation between $\hat{E}_t^{(3)}$ and $\widehat{PF}^{(3)}$. Mainly we are supposed to have:
\begin{equation}\label{eq:factor}
	\begin{split}
		D=2 & \, : \quad
		\overbrace{ \ \hat{E}_t^{(3)} \vphantom{\widehat{PF}^{(3)}} \ }^{\text{order }10} = \overbrace{ \ \bf{B}\vphantom{\widehat{PF}^{(3)}} \ }^{\text{order }8} \times  \overbrace{\  \widehat{PF}^{(3)} \ }^{\text{order }2}
		\\
		D\neq 2 & \, : \quad \overbrace{ \ \hat{E}_t^{(3)} \vphantom{\widehat{PF}^{(3)}} \ }^{\text{order }10} = \overbrace{ \ \bf{B}\vphantom{\widehat{PF}^{(3)}} \ }^{\text{order }6} \times  \overbrace{\  \widehat{PF}^{(3)} \ }^{\text{order }4}
	\end{split}		
\end{equation}
Of course knowing the explicit answer for the r.h.s factor we can manually check that it divides l.h.s operator. However, we lack the understanding for the reason to expect $\hat{E}_t^{(n)}$ to be divisible and the techniques to find the r.h.s factor independently without knowing the answer beforehand. Notice that if we want to stick to operators with polynomial coefficients, then a modification of $\eqref{eq:factor}$ is due. The operator $\bf{B}$ in both cases appears to have rational coefficients, hence after multiplication of both sides we have exactly the relation announced in the introduction:
\begin{equation}
	P\cdot \hat{E}^{(n)}_t = \hat { B}\cdot \widehat{PF}^{(n)}
\end{equation}

A convenient way to study the relation between the operator and its factors is to apply it
to a formal series $\sum\limits_n u_nt^n$  and convert into recurrence relation \cite{MMRSlarge}. For $\widehat{PF}^{(3)}$ at unequal mass and $D\neq 2$ we obtain
\be
3(n+1)^2 u_n + a^{(2)}\cdot u_{n+1} +  b^{(2)}\cdot u_{n+2}+  c^{(2)}\cdot u_{n+3}
+  d^{(2)}\cdot u_{n+4}+  e^{(2)}\cdot u_{n+5}+ (n+6)^2 f^{(0)}\cdot u_{n+6}=0
\label{rec_short}
\ee
where $a,b,c,d,e,f$ are polynomials in $n$ of degrees, shown in the superscripts. The Fourier transform  $\hat{E}^{(3)}_t$ of the coordinate space equation from A.3 of \cite{Mishnyakov:2023wpd} converts into
\begin{equation}
	\begin{split}
		(n+1)^2(n+2)^2(n+3)^2 &\Big( 
		75(n+1)^2 u_n + A^{(4)}\cdot u_{n+1} + B^{(6)}\cdot u_{n+2} + C^{(6)}\cdot u_{n+3} 		+ (n+4)^2D^{(4)}\cdot u_{n+4} + 
		\\
		&+ (n+4)^2(n+5)^2 E^{(2)}\cdot u_{n+5} + (n+4)^2(n+5)^2(n+6)^2F^{(0)}\cdot u_{n+6}
		\Big)=0
		\label{rec_long}
	\end{split}
\end{equation}

where some factors are explicitly extracted from the coefficients -- the superscripts refer to the power of $n$
which remains in the irreducible factors. The question is -- in what sense is (\ref{rec_long}) divisible by (\ref{rec_short})?
Note that in this representation both are difference equations of the same 6-th order, just coefficients
are different. This means that after division we are supposed to obtain an operator which is homogeneous in $t$. Looking at the emergent structures we can actually obtain the polynomial multiplier in the r.h.s required for the polynomiality of operators:
\begin{equation}
	P=	\Big(3 t^{2} -2\left( m_{1}^{2} + m_{2}^{2} + m_{3}^{2}\right) t - m_{1}^{4} - m_{2}^{4} - m_{3}^{4}
	+ 2 m_{1}^{2} m_{2}^{2} + 2 m_{1}^{2} m_{3}^{2} + 2 m_{2}^{2} m_{3}^{2} \Big)^{10}
	\label{corrfactor}
\end{equation}

\section{Conclusion}

The purpose of this note was to summarize the result of a technically complicated work,
reported in \cite{MMRSlarge}, in relatively simple words, emphasizing the various morals
which can be extracted already at this stage. The first takeaway is the somewhat well known fact that for generic values of parameters even the Picard-Fuchs equations are much more
sophisticated than those related to toric
Calabi-Yau varieties. They form a different family of operators and the respective family of varieties remains to be identified, The second takeaway is that the position space approach which is conceptually simpler to formulate produces however even more complicated equations in momentum space. They generically have a (much) greater order and are related to Picard-Fuchs operators is via highly non-trivial factorization of the form \eqref{factPEBPF}. This relation remain to be made conceptually clear, possibly requiring a new conceptual step. A positive example example is provided by the equal mass case where even at generic $D$ one can discover and prove general statements about the momentum space equation \emph{a la} \eqref{eq:coefrelforequallmassDnot2}.

\section*{Acknowledgements}
We are indebted for valuable discussions and comments to P. Suprun.\\
Our work is partly supported by the BASIS foundation and RFBR grants 21-51-46010-ST-a and 21-52-52004.

\providecommand{\href}[2]{#2}\begingroup\raggedright

\bibliographystyle{utphys}
\bibliography{FeynmanRef}{}

\endgroup

\end{document}